\documentclass[aps, pra, twocolumn, superscriptaddress,floatfix]{revtex4-2}
\usepackage[]{physics}
\usepackage[]{amsmath}
\usepackage{amssymb}
\usepackage{mathrsfs}
\usepackage{xcolor}
\definecolor{wandavision-red}{HTML}{F20505}
\definecolor{wandavision-green}{HTML}{067302}
\definecolor{wandavision-light-green}{HTML}{0ABF04}
\definecolor{wandavision-blue}{HTML}{0511F2}
\definecolor{wandavision-dark-blue}{HTML}{030A8C}
\makeatletter
\def\p@figure{\color{wandavision-red}}
\def\p@equation{\color{wandavision-blue}}
\makeatother

\usepackage[]{graphicx}
\graphicspath{{src/figures/}}
\usepackage[]{mathtools}
\usepackage{esint}
\usepackage[normalem]{ulem}
\usepackage{cancel}
\usepackage[]{slashed}
\usepackage[%
colorlinks=true,%
linkcolor=wandavision-blue,%
citecolor=wandavision-green,%
urlcolor=wandavision-dark-blue%
]{hyperref}
\usepackage[caption=false]{subfig}
\usepackage{cases}
\usepackage[]{comment}

\newcommand{\figref}[1]{\figurename~\ref{#1}}

\begin{document}

\title{
  A Microscopic Model Underpinning Macroscopic QED in Active Media
}

\author{Daigo Oue}
\email{daigo.oue@gmail.com}
\affiliation{RIKEN Centre for Advanced Photonics, Saitama 351-0198, Japan}
\affiliation{The Blackett Laboratory, Imperial College London, London SW7 2AZ, United Kingdom}

\date{\today}

\begin{abstract}
  We establish a microscopic foundation for macroscopic quantum electrodynamics in active media. Starting from a driven three-level system, we derive an effective two-level description with nonequilibrium stationary populations and employ a quantum central limit approach to obtain collective Langevin dynamics. This construction yields the electromagnetic susceptibility and quantum noise-current correlations from the same microscopic stationary state, providing a direct connection to the reservoir operators of conventional macroscopic quantum electrodynamics. We show that the optical response is governed by the population difference, whereas the noise correlations depend separately on the populations of the two levels. Consequently, at optical transparency, the imaginary part of the susceptibility vanishes while quantum fluctuations remain finite. We interpret the resulting noise statistics in terms of an effective reservoir occupation and a purification involving two vacuum modes, and establish a corresponding effective-temperature fluctuation–dissipation relation. Finally, we demonstrate that the residual fluctuations at transparency can induce finite heating of a nearby quantum probe.
\end{abstract}

\maketitle

\section{Introduction}
\label{sec:introduction}

Macroscopic quantum electrodynamics (MQED) provides a general framework for describing quantum electromagnetic fields in dispersive and absorptive material environments \cite{huttner1992quantization,gruner1996green,scheel2008macroscopic,philbin2010canonical}. In contrast to free-space quantum electrodynamics, the electromagnetic field in a dissipative medium does not constitute a closed system: the material degrees of freedom responsible for absorption are necessarily accompanied by fluctuations. At the macroscopic level, this structure is captured by decomposing the polarisation into a linear-response contribution and a noise polarisation,
\begin{equation}
    \hat{\vb{P}}(\vb{r},\omega)
    =
    \epsilon_0 \chi(\vb{r},\omega)
    \cdot
    \hat{\vb{E}}(\vb{r},\omega)
    +
    \hat{\vb{P}}_{\mathrm{N}}(\vb{r},\omega),
    \label{eq:intro_P_decomposition}
\end{equation}
where $\chi$ is the susceptibility of the medium and $\hat{\vb{P}}_{\mathrm{N}}$ accounts for the associated polarisation fluctuations. Throughout this work, we write $\chi=\chi'+i\chi''$, where $\chi' \equiv (\chi+\chi^\dagger)/2$ and $\chi'' \equiv (\chi-\chi^\dagger)/(2i)$ are the Hermitian and anti-Hermitian parts of the susceptibility, respectively. For a scalar susceptibility, $\chi'$ and $\chi''$ reduce to its usual real and imaginary parts. The commutation relations and correlation functions of $\hat{\vb{P}}_{\mathrm{N}}$ are constrained by the dissipative part $\chi''$ of the response through the fluctuation--dissipation relation \cite{gruner1995correlation,scheel2008macroscopic}. This construction allows the electromagnetic field to be consistently quantized in the presence of material loss while using the macroscopic response function as the central characterisation of the medium.

The MQED framework can also be extended to amplifying media, for which the dissipative part of the susceptibility can become negative over a finite frequency range \cite{matloob1997electromagnetic,raabe2008qed}. In such regions, the conventional noise-operator prescription for an absorbing medium is replaced by one involving creation rather than annihilation operators. Schematically, for a local isotropic medium,
\begin{equation}
    \hat{\vb{P}}_{\mathrm{N}}
    \propto
    \begin{cases}
        \sqrt{+\chi''}\,\hat{\vb{f}},
        & \chi''>0,\\[1mm]
        \sqrt{-\chi''}\,\hat{\vb{f}}^\dagger,
        & \chi''<0.
    \end{cases}
    \label{eq:intro_active_MQED}
\end{equation}
This prescription provides a consistent macroscopic quantization scheme for linear amplifying media and preserves the canonical commutation relations of the electromagnetic field. Physically, however, amplification is generated by a nonequilibrium microscopic state maintained by external pumping and dissipation. The relation between such microscopic driven--dissipative dynamics and the noise operators appearing in active-medium MQED is less explicit when the medium is specified directly through its macroscopic response.

Active and non-Hermitian electromagnetic environments have attracted considerable interest in quantum optics, where gain and loss can profoundly modify spontaneous emission and quantum dynamics \cite{akbarzadeh2019spontaneous,franke2021fermi,franke2022quantized,ren2021quasinormal}. Related developments have revealed fluctuation-induced instabilities and enhanced quantum friction in moving media \cite{oue2024stable,oue2025quantum}, as well as nonreciprocal quantum dynamics and lateral forces associated with chiral gain \cite{silveirinha2025chiral,oue2026fluctuation}. These results highlight the importance of a consistent quantum description of fluctuations in active environments, beyond their characterisation in terms of non-Hermitian electromagnetic response alone.

The ingredients required to establish this connection are familiar from several complementary approaches. Microscopic models of dielectric quantization explicitly introduce material and reservoir degrees of freedom and thereby provide a microscopic foundation for the noise polarisation of passive media \cite{huttner1992quantization,bechler1999quantum,suttorp2004field,philbin2010canonical}. Quantum-optical theories of open systems, on the other hand, provide a direct relation between dissipative dynamics and the corresponding quantum Langevin description of fluctuations \cite{gardiner1985input,breuer2002theory,gardiner2004quantum,walls2008quantum}. Finally, the quantum central-limit theorem provides a systematic route from microscopic quantum constituents to Gaussian bosonic collective fluctuations upon coarse-graining over a large number of microscopic constituents \cite{cushen1971quantum,goderis1989non,goderis1989central,fearnhead1999continuum,becker2021convergence}. These developments suggest a microscopic underpinning of active-medium MQED: starting from a driven open quantum system, one may determine its nonequilibrium steady state, construct its collective fluctuations, and derive both the linear response and the associated polarisation noise from the same microscopic dynamics.

\begin{figure}[tbp]
  \centering
  \includegraphics[width=\linewidth]{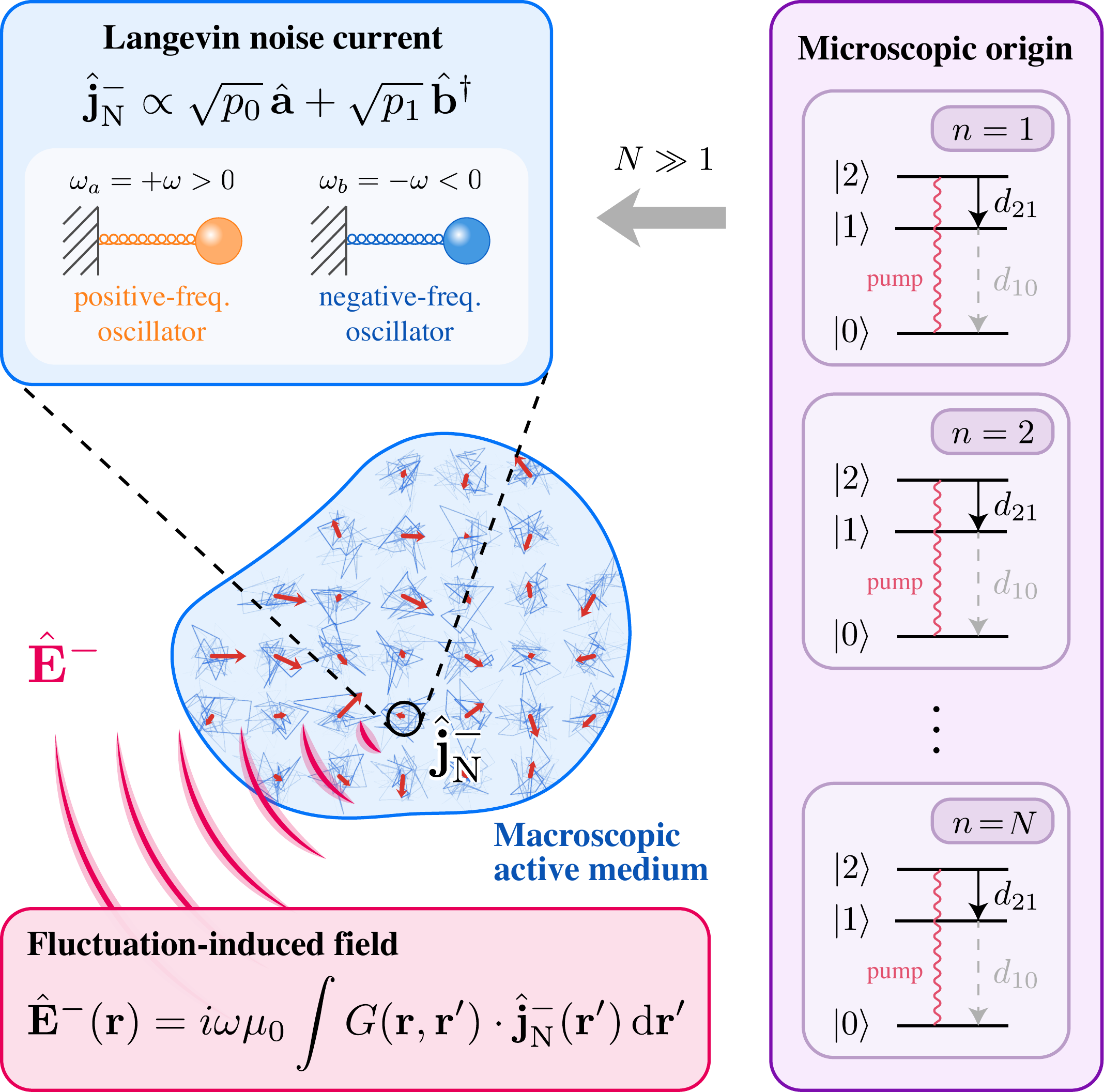}
  \caption{
    Microscopic underpinning of macroscopic QED in active media.
    An ensemble of $N\gg1$ pumped three-level systems gives rise to
    collective polarisation fluctuations, which admit an effective
    bosonic description via the quantum central limit theorem.
    At each macroscopic position, the resulting Langevin noise current $\hat{\vb{j}} _ \mathrm{N} ^ -$ is represented by positive-
    and negative-frequency oscillators ($\hat{\vb{a}}$ and $\hat{\vb{b}}$), with contributions proportional
    to $\sqrt{p_0}\,\hat{\vb{a}}$ and $\sqrt{p_1}\,\hat{\vb{b}}^\dagger$, respectively,
    where $p_0$ and $p_1$ denote the steady-state populations of the
    states $\ket{0}$ and $\ket{1}$, respectively.
    The noise current generates field fluctuations $\hat{\vb{E}} ^ -$
    through the Green tensor $G$ of the macroscopic medium.
  }
  \label{fig:fig1}
\end{figure}
In this work, we implement this program for an ensemble of coherently pumped three-level systems coupled to dissipative reservoirs (see \figref{fig:fig1}). We first eliminate the auxiliary level to obtain an effective two-level Lindblad equation whose stationary populations $p_0$ and $p_1$ are determined by the microscopic pumping and relaxation processes. We then apply the quantum central-limit theorem to the collective fluctuations of the effective two-level ensemble. In the linear-fluctuation regime, the collective lowering operator admits the Gaussian representation
\begin{equation}
    \hat{F}_{-}
    =
    \sqrt{p_0}\,\hat{a}
    +
    \sqrt{p_1}\,\hat{b}^\dagger,
    \label{eq:intro_two_channel}
\end{equation}
where $\hat{a}$ and $\hat{b}$ represent independent bosonic fluctuation channels. Using the correspondence between Lindblad and quantum Langevin descriptions, we construct the corresponding fluctuating polarisation of the coarse-grained medium. Independently, coupling the same microscopic system weakly to an electromagnetic probe yields its linear susceptibility. Consequently, both the response function and the polarisation fluctuations follow from the same nonequilibrium microscopic model rather than being introduced as independent macroscopic inputs.

The resulting structure provides a direct connection to active-medium MQED. In particular, the dissipative part of the susceptibility is controlled by the population difference,
\begin{equation}
    \chi''(\omega)
    \propto
    p_0-p_1,
    \label{eq:intro_imchi_population}
\end{equation}
whereas the two polarisation-noise correlation functions retain the populations $p_0$ and $p_1$ separately. The microscopic theory therefore naturally describes absorption, transparency, and amplification as continuous regimes of the same driven nonequilibrium system. We show that the resulting noise polarisation satisfies the standard MQED commutation relation and reproduces the generalized fluctuation--dissipation structure appropriate to an active medium \cite{matloob1997electromagnetic,raabe2008qed,oue2026macroscopic,oue2026fluctuating}. In the passive and fully inverted limits, the two fluctuation channels reduce respectively to the familiar annihilation- and creation-operator prescriptions of absorbing and amplifying MQED.

The remainder of this paper is organized as follows. In Sec.~\ref{sec:microscopic_model}, we introduce the pumped three-level model and derive its effective two-level Lindblad dynamics. In Sec.~\ref{sec:collective_fluctuations}, we construct the collective Gaussian fluctuations using the quantum central-limit theorem. In Sec.~\ref{sec:langevin}, we formulate their quantum Langevin dynamics and obtain the coarse-grained polarisation fluctuations. In Sec.~\ref{sec:mqed}, we derive the linear susceptibility and establish the connection with the noise relations of active-medium MQED. We discuss the physical interpretation of the resulting two-channel fluctuation structure in Sec.~\ref{sec:discussion}, before concluding in Sec.~\ref{sec:conclusion}.

\section{Microscopic model of an active medium}
\label{sec:microscopic_model}

\begin{figure}[htbp]
    \centering
    \includegraphics[width=.6\linewidth]{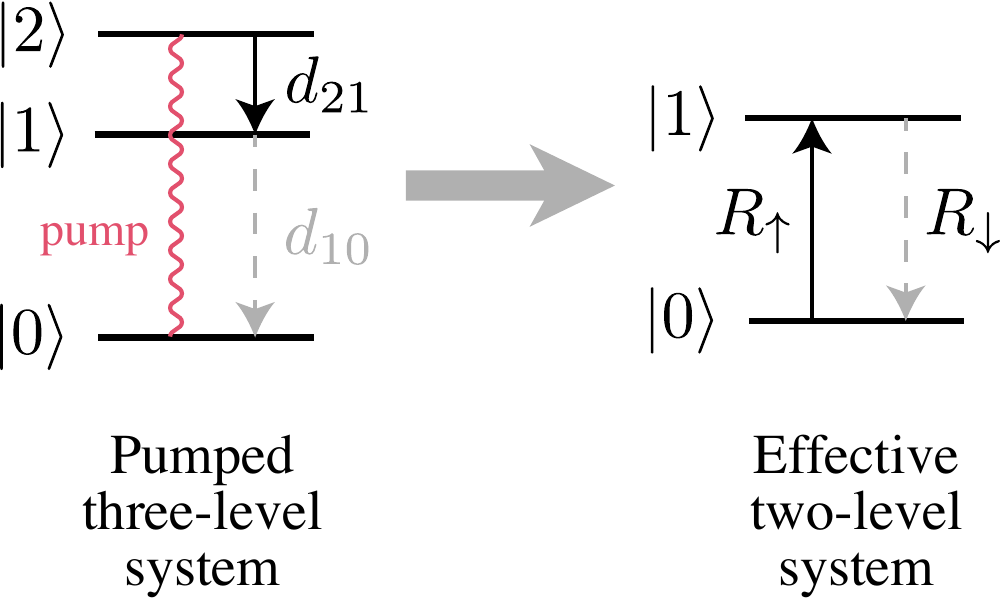}
    \caption{
        Microscopic model of an optically pumped medium.
        A driven three-level system, coupled to dissipative
        reservoirs (leading to incoherent transitions), is reduced to an effective two-level
        system (2LS) with stationary populations $p_0$
        and $p_1$.
        The effective 2LS provides the microscopic degrees
        of freedom underlying the collective fluctuations
        and electromagnetic response.
    }
    \label{fig:microscopic_model}
\end{figure}

We consider an ensemble of identical three-level systems with states $\ket{0}$, $\ket{1}$, and $\ket{2}$, as illustrated schematically in \figref{fig:microscopic_model}.
The transition between $\ket{0}$ and $\ket{2}$ is coherently driven by an external pump with Rabi frequency $g$ and detuning $\Delta$. In a frame rotating at the pump frequency, the Hamiltonian of a single three-level system may be written as
\begin{equation}
    H
    =
    \hbar\omega_{10}\ketbra{1}{1}
    +
    \hbar\Delta\ketbra{2}{2}
    +
    \frac{\hbar g}{2}
    \qty(
        \ketbra{2}{0}
        +
        \ketbra{0}{2}
    ).
    \label{eq:three_level_hamiltonian}
\end{equation}
In addition to the coherent drive, we assume that the system is coupled to thermal reservoirs representing degrees of freedom outside the discrete levels retained explicitly in the model. These reservoirs induce incoherent transitions between the levels, including the $\ket{2}\leftrightarrow\ket{1}$ and $\ket{1}\leftrightarrow\ket{0}$ transitions. We denote the corresponding downward and upward transition rates by $d_{ij}$ and $u_{ij}$, respectively. Their microscopic derivation from the reservoir coupling follows the standard Born--Markov procedure; see, for example, Ref.~\cite{campaioli2024quantum}. We will not repeat that derivation here.

We first focus on the population dynamics generated by the coherent pump and the reservoir-induced transitions. The pump generates transitions between $\ket{0}$ and the upper level $\ket{2}$. Adiabatically eliminating the rapidly decaying coherence associated with this transition yields an effective bidirectional transition between the two levels, characterised by the rate
\begin{equation}
    W
    =
    \frac{g^2\kappa_{02}}
    {2\qty(\Delta^2+\kappa_{02}^2)},
    \label{eq:pump_rate}
\end{equation}
where $\kappa_{02} = (d_{21}+u_{10})/2 +\kappa _ \phi$ denotes the decay rate of the $\ket{0}$--$\ket{2}$ coherence, where $\kappa _ \phi$ denotes the corresponding pure-dephasing rate. The derivation of Eq.~\eqref{eq:pump_rate} from the coherence dynamics is given in Appendix~\ref{appx:rate_eq}. The population of the upper level then obeys
\begin{equation}
    \dot p_2
    =
    Wp_0
    +
    u_{21}p_1
    -
    \qty(W+d_{21})p_2.
    \label{eq:p2_rate}
\end{equation}
We assume that the upper level $\ket{2}$ relaxes much faster than the two lower levels and therefore acts as an auxiliary level that can be adiabatically eliminated. Setting $\dot p_2\simeq0$ gives
\begin{equation}
    p_2
    \simeq
    \frac{Wp_0+u_{21}p_1}
    {W+d_{21}}.
    \label{eq:p2_adiabatic}
\end{equation}
We further assume that the auxiliary-level population remains negligible, $p_2\ll1$, so that the effective two-level populations satisfy $p_0+p_1\simeq1$. This condition is distinct from the adiabatic approximation $\dot p_2\simeq0$.

The resulting population dynamics within the effective two-level subspace $\{\ket{0},\ket{1}\}$ can therefore be written as
\begin{equation}
    \dot p_1
    =
    R_\uparrow p_0
    -
    R_\downarrow p_1,
    \label{eq:effective_population_rate}
\end{equation}
where
\begin{equation}
    R_\uparrow
    =
    u_{10}
    +
    \frac{Wd_{21}}{W+d_{21}},
    \qquad
    R_\downarrow
    =
    d_{10}
    +
    \frac{Wu_{21}}{W+d_{21}}.
    \label{eq:effective_rates}
\end{equation}
The origin of these rates is particularly transparent from the underlying three-level dynamics; see Appendix~\ref{appx:rate_eq} for the derivation. The second term in $R_\uparrow$ describes the pump-assisted pathway
\begin{equation}
    \ket{0}
    \xrightarrow{\,W\,}
    \ket{2}
    \xrightarrow{\,d_{21}\,}
    \ket{1},
\end{equation}
whereas the second term in $R_\downarrow$ corresponds to the reverse pathway through the upper level. The external pump can therefore enhance the effective excitation rate between $\ket{0}$ and $\ket{1}$ and, for sufficiently strong pumping, establish population inversion.

Equation~\eqref{eq:effective_population_rate} motivates an effective two-level description in terms of the operators
\begin{equation}
    \sigma_-=\ketbra{0}{1},
    \qquad
    \sigma_+=\ketbra{1}{0},
    \qquad
    \sigma_z=\ketbra{1}{1}-\ketbra{0}{0}.
\end{equation}
In addition to the population-changing processes characterised by $R_\uparrow$ and $R_\downarrow$, we allow for a general phase-relaxation channel that leaves the populations unchanged. Denoting its rate by $\gamma_\phi$, the effective dynamics is described by the Lindblad equation
\begin{equation}
    \dot\rho
    =
    -\frac{i}{\hbar}
    \comm{H_{\mathrm{eff}}}{\rho}
    +
    R_\uparrow\mathcal D _ {\sigma_+}[\rho]
    +
    R_\downarrow\mathcal D _ {\sigma_-}[\rho]
    +
    \frac{\gamma_\phi}{2}
    \mathcal D _ {\sigma_z}[\rho],
    \label{eq:effective_lindblad}
\end{equation}
where $H_{\mathrm{eff}}=\hbar\widetilde{\omega}_0\sigma_+\sigma_-$, with $\widetilde{\omega}_0$ denoting the effective transition frequency between $\ket{0}$ and $\ket{1}$, including any frequency renormalisation arising from the eliminated degrees of freedom, and $\mathcal D _ {L} [\rho]=L\rho L^\dagger-\frac{1}{2}\{L^\dagger L,\rho\}$. The corresponding transverse relaxation rate is
\begin{equation}
    \Gamma_2
    =
    \frac{R_\uparrow+R_\downarrow}{2}
    +
    \gamma_\phi.
    \label{eq:Gamma2}
\end{equation}
Here, $\Gamma_2$ governs the decay of the 2LS coherence $\langle\sigma_-\rangle$, as discussed in Section~\ref{sec:langevin}.
A systematic derivation of Eq.~\eqref{eq:effective_lindblad} using a projection-operator elimination of the upper-level degrees of freedom is given in Appendix~\ref{appx:effective_lindblad}. There we also show how the phase-relaxation term may arise microscopically from a longitudinal coupling to environmental degrees of freedom.

We now determine the nonequilibrium steady state of the effective two-level system. Setting Eq.~\eqref{eq:effective_population_rate} to zero and using $p_0+p_1=1$ gives
\begin{equation}
    p_0
    =
    \frac{R_\downarrow}
    {R_\uparrow+R_\downarrow},
    \qquad
    p_1
    =
    \frac{R_\uparrow}
    {R_\uparrow+R_\downarrow}.
    \label{eq:steady_state_populations}
\end{equation}
In the absence of a coherent drive between the effective levels, the stationary coherence vanishes, and the steady-state density operator is therefore
\begin{equation}
    \rho_{\mathrm{ss}}
    =
    p_0\ketbra{0}{0}
    +
    p_1\ketbra{1}{1}.
    \label{eq:steady_state_density}
\end{equation}
Correspondingly,
\begin{equation}
    \expval{\sigma_z}_{\mathrm{ss}}
    =
    p_1-p_0.
    \label{eq:steady_state_inversion}
\end{equation}

The populations $p_0$ and $p_1$ thus provide the microscopic parameters characterising the nonequilibrium state of the effective two-level medium. Depending on the balance between pumping and relaxation, the system can continuously pass from $p_0>p_1$ through $p_0=p_1$ to the population-inverted regime $p_1>p_0$. As shown below, these regimes will correspond respectively to absorption, transparency, and amplification in the macroscopic electromagnetic response. More importantly, the individual populations $p_0$ and $p_1$, rather than only their difference, will determine the quantum statistics of the collective polarisation fluctuations. We turn to these fluctuations in the following section.

\section{Bosonisation of collective fluctuations}
\label{sec:collective_fluctuations}

We now consider the collective fluctuations of an ensemble of the effective two-level systems (2LSs) introduced in the previous section. Consider a mesoscopic volume containing $N$ identical 2LSs, each in the nonequilibrium steady state $\rho_{\mathrm{ss}}$ given by Eq.~\eqref{eq:steady_state_density}. For a single-2LS operator $X$, we define the corresponding collective fluctuation operator as
\begin{equation}
    F^{(N)}(X)
    =
    \frac{1}{\sqrt{N}}
    \sum_{n=1}^{N}
    \qty[
        X^{(j)}
        -
        \expval{X}_{\mathrm{ss}}
    ],
    \label{eq:collective_fluctuation}
\end{equation}
where $X^{(j)}$ acts on the $j$th 2LS.

The $1/\sqrt{N}$ scaling in Eq.~\eqref{eq:collective_fluctuation} reflects the distinction between macroscopic averages and their fluctuations. For a large ensemble, quantities normalised by $N$ become self-averaging: their fluctuations vanish as $N^{-1/2}$, leaving their steady-state expectation values. The leading fluctuations about these mean values are therefore of order $\sqrt{N}$ and remain finite under the normalisation adopted in Eq.~\eqref{eq:collective_fluctuation}. The quantum central limit theorem (QCLT) states that, under appropriate clustering conditions, these collective fluctuations converge in the large-$N$ limit to Gaussian quantum variables \cite{cushen1971quantum,goderis1989non,goderis1989central,
fearnhead1999continuum,becker2021convergence}. Their covariance and commutation relations are determined entirely by the corresponding single-2LS correlations in the steady state. A brief derivation and further details of this limiting procedure are given in Appendix~\ref{appx:qclt}.

We now specialise this construction to the dipolar transition between $\ket{0}$ and $\ket{1}$. Since the steady state in Eq.~\eqref{eq:steady_state_density} is diagonal, $\expval{\sigma_-}_{\mathrm{ss}}=\expval{\sigma_+}_{\mathrm{ss}}=0$. We consider the collective fluctuation operators associated with relaxation and excitation of the 2LSs,
\begin{equation}
    F_-^{(N)}
    =
    \frac{1}{\sqrt{N}}
    \sum_{n=1}^{N}\sigma_-^{(j)},
    \qquad
    F_+^{(N)}
    =
    \qty(F_-^{(N)})^\dagger.
    \label{eq:collective_lowering}
\end{equation}
Here, we have introduced the shorthand notation $F_\pm^{(N)}\equiv F^{(N)}(\sigma_\pm)$ for the corresponding collective fluctuation operators.

For independent 2LSs in the steady state, correlations between different sites factorise. Using $\sigma_-\sigma_+=\ketbra{0}{0}$ and $\sigma_+\sigma_-=\ketbra{1}{1}$, we obtain, in the large-$N$ limit,
\begin{equation}
    \expval{F_-F_+}
    =
    p_0,
    \qquad
    \expval{F_+F_-}
    =
    p_1.
    \label{eq:collective_correlations}
\end{equation}
Here and below, $F_\pm$ denote the limiting Gaussian fluctuation operators associated with $F_\pm^{(N)}$ as $N\rightarrow\infty$.

The commutation relation follows in the same way. Since operators belonging to different 2LSs commute,
\begin{equation}
    \comm{F_-^{(N)}}{F_+^{(N)}}
    =
    \frac{1}{N}
    \sum_{n=1}^{N}
    \comm{\sigma_-^{(j)}}{\sigma_+^{(j)}}.
    \label{eq:finite_N_commutator}
\end{equation}
The quantity on the right-hand side is a macroscopic average and therefore becomes self-averaging in the large-$N$ limit. Using $\comm{\sigma_-}{\sigma_+}=\ketbra{0}{0}-\ketbra{1}{1}$, we obtain
\begin{equation}
    \comm{F_-}{F_+}
    =
    p_0-p_1.
    \label{eq:collective_commutator}
\end{equation}
Equations~\eqref{eq:collective_correlations} and \eqref{eq:collective_commutator} characterise the Gaussian fluctuation algebra inherited from the microscopic steady state.

A convenient bosonic representation of this limiting fluctuation algebra can be constructed using two independent canonical bosonic modes, $a$ and $b$, satisfying
\begin{equation}
    \comm{a}{a^\dagger}
    =
    \comm{b}{b^\dagger}
    =
    1,
    \qquad
    \comm{a}{b}
    =
    \comm{a}{b^\dagger}
    =
    0.
    \label{eq:canonical_bosonic_modes}
\end{equation}
Taking these modes in their vacuum states, the collective fluctuation operator can be represented as
\begin{equation}
    F_-
    =
    \sqrt{p_0}\,a
    +
    \sqrt{p_1}\,b^\dagger.
    \label{eq:bosonic_representation}
\end{equation}
Indeed, Eq.~\eqref{eq:bosonic_representation} immediately gives $\expval{F_-F_+}=p_0$ and $\expval{F_+F_-}=p_1$, together with $\comm{F_-}{F_+}=p_0-p_1$, and therefore reproduces both the second moments and the commutation relation obtained from the microscopic ensemble.

The collective fluctuation of the microscopic dipoles is thus represented, in the large-$N$ limit, by a linear combination of a bosonic annihilation operator and an independent bosonic creation operator. Importantly, their respective weights are fixed directly by the steady-state populations as $\sqrt{p_0}$ and $\sqrt{p_1}$. Equation~\eqref{eq:bosonic_representation} provides the microscopic starting point for the Langevin description of the polarisation fluctuations developed in the following section.

\section{Langevin dynamics of collective fluctuations}
\label{sec:langevin}

Having established the bosonic representation of the collective fluctuations in Sec.~\ref{sec:collective_fluctuations}, we now turn to their dynamics. We assume that the 2LSs evolve independently under the effective Lindblad dynamics introduced in Sec.~\ref{sec:microscopic_model}, and that their coupling to the reservoirs is local and Markovian. Under these assumptions, the collective fluctuations become Gaussian in the large-$N$ limit, as established in the previous section. Their Markovian dynamics can then be represented by an equivalent quantum Langevin equation, following the standard correspondence between master equations and quantum stochastic dynamics~\cite{gardiner1985input}.

The adjoint of the effective Lindblad generator gives the equation of motion for the single-2LS coherence,
\begin{equation}
  \qty(
  \frac{d}{dt}
  +i\widetilde{\omega}_0+\Gamma_2
  )
  \expval{\sigma_-}
  =
  0,
\end{equation}
where $\Gamma_2$ is the transverse relaxation rate defined in Eq.~\eqref{eq:Gamma2}. The corresponding collective fluctuation dynamics can be represented by the Langevin equation
\begin{equation}
    \dot F_-(t)
    =
    -\qty(i\widetilde{\omega}_0+\Gamma_2)F_-(t)
    +
    \xi(t),
    \label{eq:collective_langevin}
\end{equation}
where $\xi(t)$ is a quantum noise operator accounting for the fluctuations associated with the eliminated reservoir degrees of freedom.

A convenient representation of the Langevin force is obtained by introducing two independent bosonic vacuum input fields, $a_{\mathrm{in}}(t)$ and $b_{\mathrm{in}}(t)$, satisfying
\begin{equation}
    \comm{a_{\mathrm{in}}(t)}{a_{\mathrm{in}}^\dagger(t')}
    =
    \comm{b_{\mathrm{in}}(t)}{b_{\mathrm{in}}^\dagger(t')}
    =
    \delta(t-t'),
    \label{eq:input_commutators}
\end{equation}
with vanishing cross-commutators. The Langevin force can then be written as
\begin{equation}
    \xi(t)
    =
    \sqrt{2\Gamma_2}
    \qty[
        \sqrt{p_0}\,a_{\mathrm{in}}(t)
        +
        \sqrt{p_1}\,b_{\mathrm{in}}^\dagger(t)
    ].
    \label{eq:langevin_noise_representation}
\end{equation}
The weights of the two input fields are determined by the steady-state populations, in direct correspondence with the bosonic representation in Eq.~\eqref{eq:bosonic_representation}. A construction of this Langevin representation from independent damped bosonic modes, together with a verification of its consistency with the effective Lindblad dynamics, is given in Appendix~\ref{appx:langevin}.

For vacuum input fields, Eq.~\eqref{eq:langevin_noise_representation} gives
\begin{equation}
    \begin{aligned}
        \expval{\xi(t)\xi^\dagger(t')}
        &=
        2\Gamma_2p_0\,\delta(t-t'),\\
        \expval{\xi^\dagger(t)\xi(t')}
        &=
        2\Gamma_2p_1\,\delta(t-t').
    \end{aligned}
    \label{eq:langevin_noise_correlations}
\end{equation}
Consequently,
\begin{equation}
    \comm{\xi(t)}{\xi^\dagger(t')}
    =
    2\Gamma_2(p_0-p_1)\delta(t-t').
    \label{eq:langevin_noise_commutator}
\end{equation}
These noise correlations ensure that the stationary solution of Eq.~\eqref{eq:collective_langevin} reproduces the collective correlations obtained in Eq.~\eqref{eq:collective_correlations}, while preserving the commutation relation in Eq.~\eqref{eq:collective_commutator}.

We next extend the description to a spatially continuous medium. We divide the medium into mesoscopic cells of volume $\Delta V$, each containing $N=n\Delta V$ identical 2LSs, where $n$ is their number density. Each cell is assumed to contain sufficiently many 2LSs for the Gaussian fluctuation description to apply, while remaining small compared with the characteristic length scales over which the macroscopic electromagnetic fields vary. Introducing the continuum-normalised fluctuation field
\begin{equation}
    f(\vb r,t)
    =
    \frac{F_-(\vb r,t)}{\sqrt{\Delta V}},
    \label{eq:continuum_fluctuation}
\end{equation}
the collective Langevin equation becomes
\begin{equation}
    \qty(
    \pdv{t}
    +i\widetilde{\omega}_0+\Gamma_2)
    f(\vb r,t)
    =
    \zeta(\vb r,t),
    \label{eq:continuum_langevin}
\end{equation}
where $\zeta(\vb r,t)=\xi(\vb r,t)/\sqrt{\Delta V}$ is the continuum-normalised Langevin force.
Note that we may omit the arguments for brevity in the following where appropriate.
The assumption of independent 2LSs and local reservoirs gives
\begin{equation}
    \begin{aligned}
        \expval{
            \zeta(\vb r,t)
            \zeta^\dagger(\vb r',t')
        }
        &=
        2\Gamma_2p_0
        \delta _ {\vb r,\vb r'}
        \delta _ {t,t'},\\
        \expval{
            \zeta^\dagger(\vb r,t)
            \zeta(\vb r',t')
        }
        &=
        2\Gamma_2p_1
        \delta _ {\vb r,\vb r'}
        \delta _ {t,t'}.
    \end{aligned}
    \label{eq:continuum_noise_correlations}
\end{equation}
Here and throughout, we use the shorthand $\delta_{x,x'}:=\delta(x-x')$ for Dirac delta distributions, where $x$ may denote a time, frequency, or position vector.

Finally, we relate the collective fluctuations to the microscopic electric dipoles. Let $\vb d$ denote the transition dipole moment associated with the $\ket{0}\leftrightarrow\ket{1}$ transition, assumed to have a fixed orientation throughout the medium. The positive-frequency polarisation operator is
\begin{equation}
    \hat{\vb P}^{-}(\vb r,t)
    =
    \vb d\sqrt{n}\,f(\vb r,t).
    \label{eq:microscopic_polarisation}
\end{equation}
Although the polarisation is treated as a vector field, the fixed transition dipole orientation restricts its fluctuations to a single polarisation channel. The extension to an isotropic medium with randomly oriented transition dipoles, in which microscopic fluctuation channels are systematically combined into three effective bosonic noise channels while preserving their correlations, is presented in Appendix~\ref{appx:isotropic}. Equation~\eqref{eq:continuum_langevin} therefore gives
\begin{equation}
  \qty(
  \pdv{t}
  +i\widetilde{\omega}_0+\Gamma_2)
  \hat{\vb P}^{-}(\vb r,t)
  =
  \hat{\vb\Xi}(\vb r,t),
  \label{eq:polarisation_langevin}
\end{equation}
where $\hat{\vb\Xi}(\vb r,t)=\vb d\sqrt{n}\,\zeta(\vb r,t)$.

We have thus obtained a Langevin description of the collective polarisation fluctuations whose dynamics and noise correlations are fixed by the microscopic nonequilibrium steady state. In the following section, we introduce the coupling to the electromagnetic field and show how the macroscopic susceptibility and the associated noise polarisation emerge from this description.

\section{Microscopic foundations of active-medium macroscopic QED}
\label{sec:mqed}

Having established the Langevin dynamics of the collective polarisation fluctuations in Sec.~\ref{sec:langevin}, we now introduce their coupling to the electromagnetic field. Our aim is to derive the macroscopic susceptibility and the associated noise polarisation from the same microscopic nonequilibrium steady state, thereby establishing their mutual consistency.

We consider the electric-dipole interaction
\begin{equation}
    \hat H_{\mathrm{int}}
    =
    -\sum_j
    \qty(
        \vb d\,\sigma_-^{(j)}
        +
        \vb d^*\,\sigma_+^{(j)}
    )
    \cdot\hat{\vb E}(\vb r_j),
    \label{eq:dipole_interaction}
\end{equation}
where $\vb d$ is the generally complex transition dipole moment introduced in the preceding section. We work within the rotating-wave approximation and consider the linear response to a weak electromagnetic probe. In this regime, the population difference can be replaced by its stationary value $p_0-p_1$, while the reservoir-induced fluctuations remain governed by the Langevin force derived in Sec.~\ref{sec:langevin}.

The equation of motion for the positive-frequency polarisation becomes
\begin{equation}
  \qty(
  \pdv{t}
  +i\widetilde{\omega}_0+\Gamma_2
  )
  \hat{\vb P}^{-}
  =
  \frac{i n(p_0-p_1)}{\hbar}
  \vb d\,\vb d^*\cdot\hat{\vb E}^{-}
  +
  \hat{\vb\Xi}.
  \label{eq:driven_polarisation_langevin}
\end{equation}
Here, the dyadic $\vb d\,\vb d^*$ acts on the electric field, and $\hat{\vb\Xi}$ is the polarisation Langevin force introduced in Eq.~\eqref{eq:polarisation_langevin}.

We adopt the Fourier-transform convention
\begin{equation}
  \widetilde X(\omega)
  =
  \int_{-\infty}^{\infty}
  X(t)
  e^{i\omega t}
  \dd{t},
  \quad
  X(t)
  =
  \int_{-\infty}^{\infty}
  \widetilde X(\omega)
  e^{-i\omega t}
  \frac{\dd{\omega}}{2\pi}.
  \label{eq:fourier_convention}
\end{equation}
Following the standard convention of MQED, the factor $1/(2\pi)$ is absorbed into the definition of the frequency-resolved positive-frequency operator:
\begin{equation}
    \hat X^{-}(t)
    =
    \int_0^\infty
    \hat X^{-}(\omega)
    e^{-i\omega t}
    \dd{\omega},
    \quad
    \hat X^{-}(\omega)
    =
    \frac{\widetilde X^{-}(\omega)}{2\pi}.
    \label{eq:mqed_fourier_convention}
\end{equation}
The corresponding negative-frequency operator is defined by Hermitian conjugation,
$\hat X^{+}(\omega)=[\hat X^{-}(\omega)]^\dagger$.

Fourier transforming Eq.~\eqref{eq:driven_polarisation_langevin} gives
\begin{equation}
    \hat{\vb P}^{-}(\vb r,\omega)
    =
    \epsilon_0\boldsymbol{\chi}(\omega)
    \cdot\hat{\vb E}^{-}(\vb r,\omega)
    +
    \hat{\vb P}_{\mathrm N}^{-}(\vb r,\omega),
    \label{eq:polarisation_constitutive_relation}
\end{equation}
where the susceptibility tensor is
\begin{equation}
    \chi_{ij}(\omega)
    =
    \frac{n d_i d_j^*}{\epsilon_0\hbar}
    \frac{p_0-p_1}
    {\widetilde{\omega}_0-\omega-i\Gamma_2},
    \label{eq:microscopic_susceptibility}
\end{equation}
and the noise polarisation is
\begin{equation}
    \hat{\vb P}_{\mathrm N}^{-}(\vb r,\omega)
    =
    \frac{\hat{\vb\Xi}^{-}(\vb r,\omega)}
    {\Gamma_2+i(\widetilde{\omega}_0-\omega)}.
    \label{eq:noise_polarisation}
\end{equation}
Thus, the response and the noise arise from the same effective 2LS dynamics, rather than being specified independently. Although Eq.~\eqref{eq:noise_polarisation} differs in appearance from the conventional MQED representation, the two are equivalent away from transparency. We exhibit this correspondence below in the absorptive regime; the corresponding amplifying representation and the operator normalisation are detailed in Appendix~\ref{appx:mqed_consistency}. The transparency limit is discussed in Sec.~\ref{sec:effective_temperature}.

The dissipative part of the susceptibility is defined by the anti-Hermitian part of the response tensor,
\begin{equation}
    \boldsymbol{\chi}''(\omega)
    =
    \frac{
        \boldsymbol{\chi}(\omega)
        -
        \boldsymbol{\chi}^\dagger(\omega)
    }{2i}.
    \label{eq:chi_antihermitian}
\end{equation}
For the susceptibility in Eq.~\eqref{eq:microscopic_susceptibility}, this gives
\begin{equation}
    \chi''_{ij}(\omega)
    =
    \frac{n d_i d_j^*}{\epsilon_0\hbar}
    \frac{(p_0-p_1)\Gamma_2}
    {(\widetilde{\omega}_0-\omega)^2+\Gamma_2^2}.
    \label{eq:microscopic_chi_imaginary}
\end{equation}
The sign of $\boldsymbol{\chi}''$ is therefore controlled by the stationary population difference. The medium is absorptive for $p_0>p_1$, amplifying for $p_1>p_0$, and transparent in linear response when $p_0=p_1$.

For the fixed transition-dipole orientation considered here, the dissipative susceptibility in Eq.~\eqref{eq:microscopic_chi_imaginary} has rank at most one. Defining the normalised dipole vector $\vb u=\vb d/|\vb d|$, we may write
\begin{equation}
    \boldsymbol{\chi}''(\omega)
    =
    \chi''_{\parallel}(\omega)\,\vb u\vb u^\dagger,
    \quad
    \chi''_{\parallel}(\omega)
    =
    \frac{n|\vb d|^2}{\epsilon_0\hbar}
    \frac{(p_0-p_1)\Gamma_2}
    {(\widetilde{\omega}_0-\omega)^2+\Gamma_2^2}.
    \label{eq:mqed_rank_one}
\end{equation}
In the absorptive regime $p_0>p_1$, the positive-semidefinite matrix square root is
$\sqrt{\boldsymbol{\chi}''}=\sqrt{\chi''_{\parallel}}\,\vb u\vb u^\dagger$.
After absorbing the frequency-dependent phase of the Langevin propagator into the input operators (see Appendix~\ref{appx:mqed_consistency}), the noise polarisation can be expressed as
\begin{equation}
    \hat{\vb P}_{\mathrm N}^{-}(\vb r,\omega)
    =
    \sqrt{\frac{\hbar\epsilon_0}{\pi}}\,
    \sqrt{\boldsymbol{\chi}''(\omega)}
    \cdot\hat{\vb f}(\vb r,\omega),
    \label{eq:mqed_standard_noise}
\end{equation}
where the operator-valued vector is
\begin{equation}
    \hat{\vb f}(\vb r,\omega)
    =
    \vb u\left[
        \sqrt{\frac{p_0}{p_0-p_1}}\,\hat a(\vb r,\omega)
        +
        \sqrt{\frac{p_1}{p_0-p_1}}\,\hat b^\dagger(\vb r,\omega)
    \right].
    \label{eq:mqed_effective_noise}
\end{equation}
Here $\hat a$ and $\hat b$ denote independent, delta-normalised vacuum bosonic fields, with the Lorentzian phase already incorporated into their definitions. Their precise relation to the time-domain Langevin inputs, including Fourier normalisation, is given in the same appendix. Equation~\eqref{eq:mqed_standard_noise} has the conventional MQED structure, but only the dipole-aligned polarisation channel is present; no independent transverse noise components need be introduced.

The projected operator $\hat{\vb f}=\vb u\hat c$ satisfies $[\hat f_i(\vb r,\omega),\hat f_j^\dagger(\vb r',\omega')]=u_i u_j^*\delta_{\vb r,\vb r'}\delta_{\omega,\omega'}$; its transverse components are not independent canonical noise channels. The corresponding effective reservoir occupation is considered in Sec.~\ref{sec:effective_reservoir}.

We now examine the noise correlations. Using the local Markovian correlations obtained in Sec.~\ref{sec:langevin}, together with the Fourier convention in Eq.~\eqref{eq:mqed_fourier_convention}, we obtain
\begin{subequations}
\label{eq:noise_polarisation_correlations}
\begin{align}
    \expval{
        \hat P_{\mathrm N,i}^{-}(\vb r,\omega)
        \hat P_{\mathrm N,j}^{+}(\vb r',\omega')
    }
    &=
    \frac{\hbar\epsilon _ 0}{\pi}
    \frac{p_0}{p _ 0 - p _ 1}
    \chi'' _ {ij}(\omega)
    \delta _ {\vb{r},\vb{r}'}
    \delta _ {\omega,\omega'}.
    \\
    \expval{
        \hat P_{\mathrm N,j}^{+}(\vb r',\omega')
        \hat P_{\mathrm N,i}^{-}(\vb r,\omega)
    }
    &=
    \frac{\hbar\epsilon _ 0}{\pi}
    \frac{p_1}{p _ 0 - p _ 1}
    \chi'' _ {ij}(\omega)
    \delta _ {\vb{r},\vb{r}'}
    \delta _ {\omega,\omega'}.
\end{align}
\end{subequations}
These correlations are weighted by the lower- and upper-state populations, respectively. Note that $\boldsymbol{\chi}''/(p_0-p_1)$ has a finite limit at $p_0=p_1$, so both noise correlations remain finite at transparency; we return to this point in Sec.~\ref{sec:discussion}.

To connect with the standard MQED formulation, we introduce the noise-current operator
\begin{equation}
    \hat{\vb j}_{\mathrm N}^{-}(\vb r,\omega)
    =
    -i\omega\hat{\vb P}_{\mathrm N}^{-}(\vb r,\omega).
    \label{eq:microscopic_noise_current}
\end{equation}
The noise-current commutator then becomes
\begin{equation}
  \comm{
    \hat j_{\mathrm N,i}^{-}(\vb r,\omega)
  }{
    \hat j_{\mathrm N,j}^{+}(\vb r',\omega')
  }
  =
  \frac{\omega^2}{c^2}
  \frac{\hbar}{\pi\mu _ 0}
  \chi''_{ij}(\omega)
  \delta _ {\vb r,\vb r'}
  \delta _ {\omega,\omega'}.
  \label{eq:mqed_noise_commutator}
\end{equation}
This relation establishes the consistency between the microscopic noise dynamics and the dissipative part of the macroscopic electromagnetic response.

The corresponding Maxwell equation takes the familiar macroscopic form
\begin{equation}
    \nabla\times\nabla\times
    \hat{\vb E}^{-}(\vb r,\omega)
    -
    \frac{\omega^2}{c^2}
    \boldsymbol{\epsilon}(\omega)
    \cdot\hat{\vb E}^{-}(\vb r,\omega)
    =
    i\omega\mu_0
    \hat{\vb j}_{\mathrm N}^{-}(\vb r,\omega),
    \label{eq:mqed_wave_equation}
\end{equation}
where $\boldsymbol{\epsilon}(\omega)=\mathbf I+\boldsymbol{\chi}(\omega)$. Its solution can be expressed in terms of the electromagnetic Green tensor, with the volume element abbreviated as $\dd{\vb r}:=\dd{x}\,\dd{y}\,\dd{z}$:
\begin{equation}
    \hat{\vb E}^{-}(\vb r,\omega)
    =
    i\omega\mu_0
    \int
    \mathbf G(\vb r,\vb r',\omega)
    \cdot
    \hat{\vb j}_{\mathrm N}^{-}(\vb r',\omega)
    \dd{\vb{r}'},
    \label{eq:mqed_green_solution}
\end{equation}
where any independently specified homogeneous-field contribution is understood to be treated separately.

In conventional macroscopic QED, the noise-current operator is constructed from bosonic reservoir operators, while the reservoir state and its nonequilibrium correlations must generally be specified separately~\cite{gruner1995correlation,gruner1996green,scheel2008macroscopic}. Here, both the response, governed by $p_0-p_1$, and the noise statistics, governed by $p_0$ and $p_1$ individually, follow from the same microscopic steady state. Their physical implications, particularly at transparency, are discussed in Sec.~\ref{sec:discussion}. An extension to randomly oriented 2LSs, in which independent microscopic fluctuations are coarse-grained into three effective polarisation noise channels, is outlined in Appendix~\ref{appx:isotropic}.

\section{Discussion}
\label{sec:discussion}

The microscopic construction developed above provides both the electromagnetic response and the statistics of the accompanying fluctuations. We now discuss the purification and occupation of the effective reservoir, examine the effective fluctuation--dissipation relation, and consider the response of a quantum probe placed near the pumped medium.

\subsection{Effective reservoir occupation and its microscopic origin}
\label{sec:effective_reservoir}

The reservoir representation established in Sec.~\ref{sec:mqed} makes it possible to interpret the microscopic origin of the effective noise statistics. In the absorptive regime, the dipole-aligned operator $\hat c$ defined through Eq.~\eqref{eq:mqed_effective_noise} is a canonical bosonic operator, although it is built from two independent vacuum fields. Writing $\hat{\vb f}=\vb u\hat c$, the effective scalar operator obeys
\begin{equation}
\begin{aligned}
    \comm{\hat c(\vb r,\omega)}{\hat c^\dagger(\vb r',\omega')}
    &=\delta_{\vb r,\vb r'}\delta_{\omega,\omega'},\\
    \expval{\hat c^\dagger(\vb r,\omega)\hat c(\vb r',\omega')}
    &=n_{\mathrm{eff}}\delta_{\vb r,\vb r'}\delta_{\omega,\omega'}.
\end{aligned}
    \label{eq:discussion_effective_occupation}
\end{equation}
with
\begin{equation}
    n_{\mathrm{eff}}
    =
    \frac{p_1}{p_0-p_1},
    \quad (p_0>p_1).
    \label{eq:effective_reservoir_occupation}
\end{equation}
Thus the effective bosonic reservoir is generally populated even though the two underlying input fields are in their vacuum states. In the language of Gaussian-state purification, the two-vacuum-mode construction is a dilation of a populated effective reservoir: the creation-operator contribution of the second vacuum field produces the nonzero occupation of $\hat c$. The orthogonal Bogoliubov partner is absent from the polarisation operator but is needed to realise the transformation on the full two-mode Hilbert space. Consequently, the effective occupation is not an additional phenomenological assumption: it is fixed by the same microscopic populations that determine the susceptibility.

For $p_1>p_0$, the susceptibility is amplifying and the corresponding conventional representation uses a creation operator multiplied by $\sqrt{-\boldsymbol{\chi}''}$. The operator transformation is analogous, but its normalisation changes sign. At transparency, $p_0=p_1$, the single-canonical-mode representation becomes singular even though the original two-vacuum-mode noise polarisation remains regular. These distinctions are made explicit in Appendix~\ref{appx:mqed_consistency}; their physical consequence is discussed next.

\subsection{Effective-temperature fluctuation--dissipation relation}
\label{sec:effective_temperature}

\begin{figure*}[tbp]
    \centering
    \includegraphics[width=\linewidth]{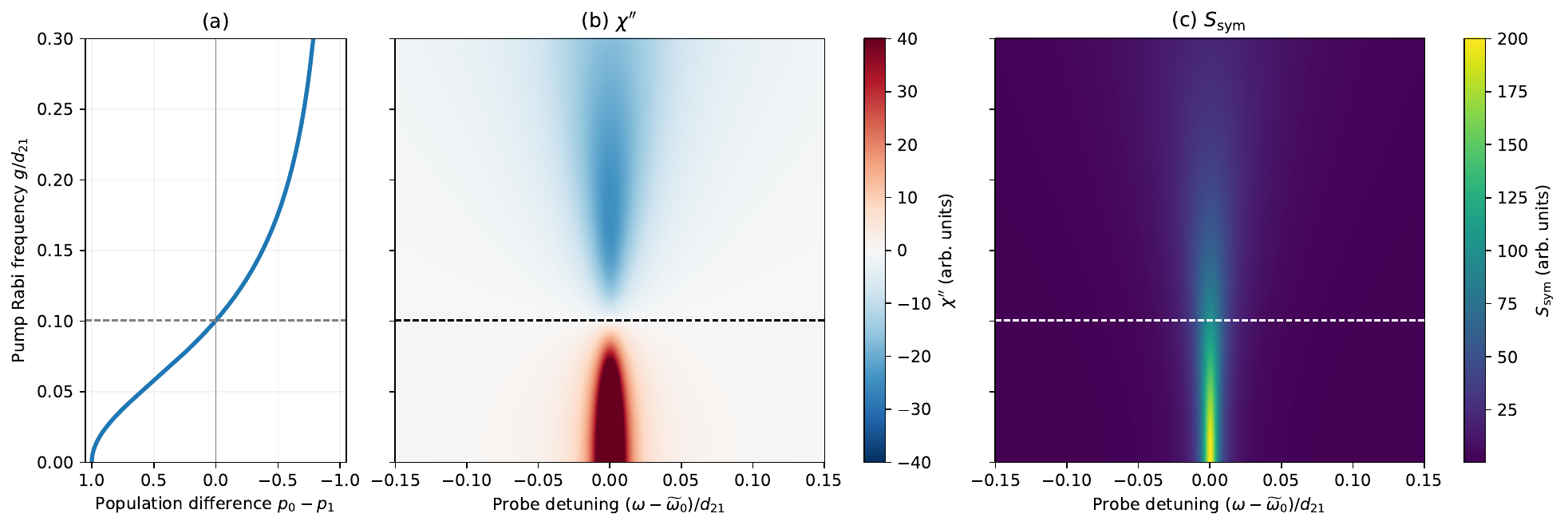}
    \caption{
      Optical response and quantum fluctuations across transparency for a resonantly pumped medium with $d_{10}/d_{21}=0.01$. The dashed line indicates the transparency point ($g/d_{21}\approx0.1$).
      (a) Stationary population difference $p_0-p_1$ as a function of pump strength.
      (b) Imaginary (anti-Hermitian) part $\chi''$ of the susceptibility, showing the transition from absorption to amplification.
      (c) Corresponding symmetrised noise spectrum $S_{\mathrm{sym}}$, which remains finite at transparency despite the vanishing dissipative susceptibility. The spectra in (b) and (c) are shown in arbitrary units.
      Note that the delta functions enforcing spatial locality and frequency conservation are omitted, and only the corresponding spectral density is plotted in (c).
      The parameters are chosen as $d_{10}/d_{21}=0.01$. For simplicity, we assume resonant pumping ($\Delta=0$) and neglect incoherent upward transitions ($u_{10}=u_{21}=0$) and pure dephasing ($\gamma_\phi=\kappa_\phi=0$), so that $\kappa_{02}=d_{21}/2$.
    }
    \label{fig:response_fluctuations}
\end{figure*}

The ratio of the two noise correlations in Eq.~\eqref{eq:noise_polarisation_correlations} is set by the stationary population ratio $p_1/p_0$. It is therefore useful to introduce a transition-specific effective inverse temperature through
\begin{equation}
    \frac{p_1}{p_0}
    =
    e^{-\beta_{\mathrm{eff}}\hbar\widetilde{\omega}_0},
    \quad
    \beta_{\mathrm{eff}}
    =
    \frac{1}{\hbar\widetilde{\omega}_0}
    \ln\frac{p_0}{p_1}.
    \label{eq:effective_inverse_temperature}
\end{equation}
Here $\beta_{\mathrm{eff}}=1/(k_{\mathrm B}T_{\mathrm{eff}})$, when an effective temperature is assigned. This definition characterises the population ratio of the selected 2LS transition; it does not imply global thermal equilibrium of the pumped medium. In particular, the frequency dependence of the response remains centred on $\widetilde{\omega}_0$, while the population ratio is fixed by the driven stationary state.

Combining Eqs.~\eqref{eq:noise_polarisation_correlations} and \eqref{eq:effective_inverse_temperature}, the symmetrised noise spectrum can be written in the fluctuation--dissipation-like form, where $\acomm{A}{B}=AB+BA$,
\begin{equation}
\begin{aligned}
    S _ {ij} ^ \mathrm{sym} 
    &= \expval{\acomm{
        \hat P_{\mathrm N,i}^{-}(\vb r,\omega)
      }{
        \hat P_{\mathrm N,j}^{+}(\vb r',\omega')
    }}
    \\
    &=\frac{\hbar\epsilon_0}{\pi}
    \coth\left(\frac{\beta_{\mathrm{eff}}\hbar\widetilde{\omega}_0}{2}\right)
    \chi''_{ij}(\omega)
    \delta_{\vb r,\vb r'}\delta_{\omega,\omega'}.
\end{aligned}
    \label{eq:effective_fdt}
\end{equation}
The tensor ordering in this expression follows from the Hermitian dyadic structure of $\boldsymbol{\chi}''$. Although population inversion reverses the sign of $\boldsymbol{\chi}''$, the effective-temperature factor changes sign simultaneously, ensuring that the symmetrised noise correlation remains positive semidefinite. The relation resembles the equilibrium fluctuation--dissipation theorem, but the hyperbolic cotangent is evaluated at the 2LS transition energy rather than at the observation frequency $\hbar\omega$; it is therefore not a claim of a frequency-independent thermal bath satisfying equilibrium KMS relations at every frequency.

In the absorptive regime $\beta_{\mathrm{eff}}>0$, whereas population inversion gives $\beta_{\mathrm{eff}}<0$ and $\boldsymbol{\chi}''<0$. Their product in Eq.~\eqref{eq:effective_fdt} remains positive semidefinite. At transparency $p_0=p_1$, the linear dissipative response vanishes, $\boldsymbol{\chi}''=0$, and $\beta_{\mathrm{eff}}\to0$. Nevertheless, the product in Eq.~\eqref{eq:effective_fdt} has a finite limit because $\boldsymbol{\chi}''$ is proportional to $p_0-p_1$. The individual normally ordered and antinormally ordered noise correlations in Eq.~\eqref{eq:noise_polarisation_correlations} likewise remain finite. Hence vanishing linear absorption does not imply vanishing electromagnetic fluctuations. The microscopic two-channel representation makes this distinction manifest without invoking a singular effective occupation.

Figure~\ref{fig:response_fluctuations} illustrates these features for a resonantly pumped medium with $d_{10}/d_{21}=0.01$, vanishing incoherent upward rates, and no pure dephasing. As the pump strength increases, the population difference changes sign, driving the transition from absorption to amplification. In contrast, the symmetrised noise spectrum remains finite and continuous across transparency, demonstrating that the cancellation of the dissipative response does not entail the suppression of quantum fluctuations.

\subsection{Heating of a quantum probe near a pumped medium}
\label{sec:quantum_probe}

\begin{figure}[tbp]
    \centering
    \includegraphics[width=.6\linewidth]{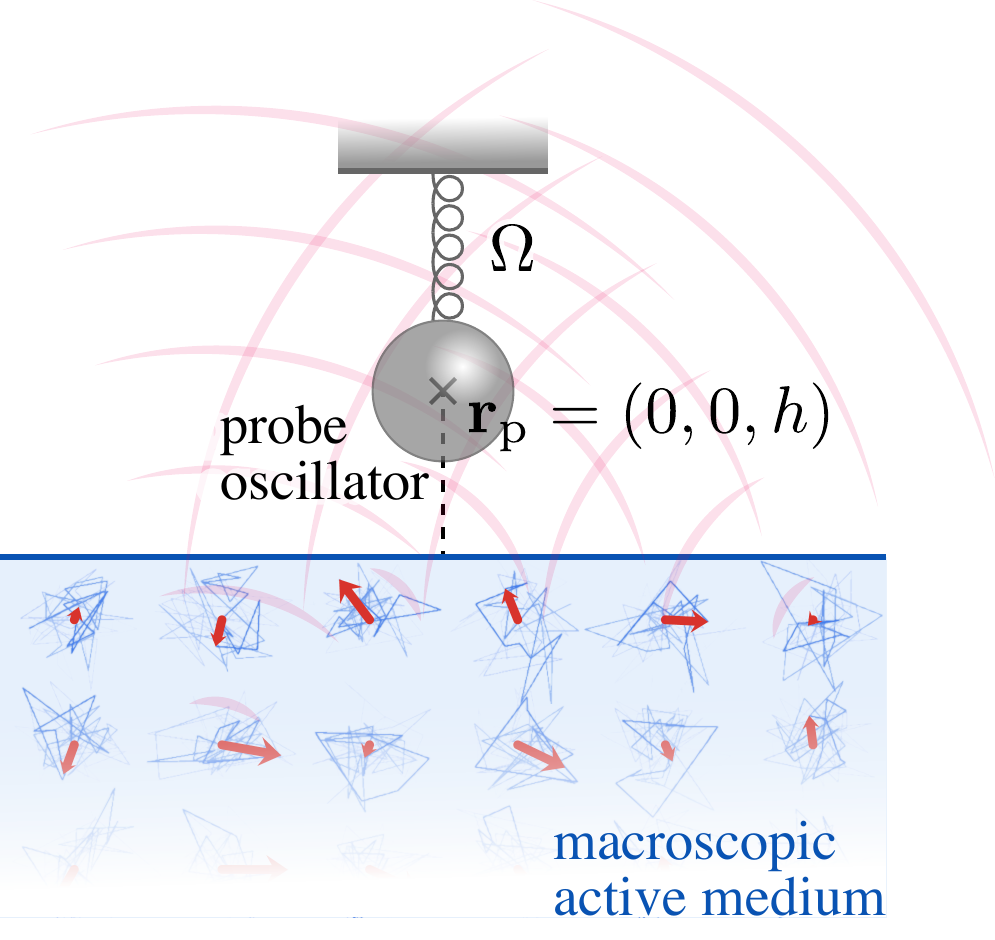}
    \caption{
        Quantum probe of fluctuations near an optically
        pumped medium.
        A harmonic oscillator of frequency $\Omega$ is
        placed at a height $h$ above a pumped half-space
        occupying $z<0$.
        Electromagnetic fluctuations generated by the
        medium couple to the probe and induce excitation
        and relaxation processes.
        At transparency, the medium-induced excitation
        rate remains finite despite the vanishing
        imaginary part of the susceptibility.
    }
    \label{fig:quantum_probe}
\end{figure}

As an operational consequence, consider a harmonic-oscillator probe of frequency $\Omega>0$ at $\vb r_{\mathrm p}=(0,0,h)$ above a pumped half-space occupying $z<0$ (see \figref{fig:quantum_probe}). 

Let $\hat c_{\mathrm p}$ annihilate a probe excitation (the corresponding Hamiltonian is $\hat{H} _ \mathrm{p} = \hbar\Omega\hat{c} _ \mathrm{p} ^ \dagger \hat{c} _ \mathrm{p}$), and let its transition dipole be $\vb d_{\mathrm p}$. In the electric-dipole and rotating-wave approximations, the interaction with the local field is
\begin{equation}
    \hat H_{\mathrm{pf}}
    =
    -\hat c_{\mathrm p}^\dagger
    \vb d_{\mathrm p}^{\,*}\cdot
    \hat{\vb E}^{-}(\vb r_{\mathrm p})
    -
    \hat c_{\mathrm p}
    \vb d_{\mathrm p}\cdot
    \hat{\vb E}^{+}(\vb r_{\mathrm p}).
    \label{eq:probe_interaction}
\end{equation}
We assume weak probe--field coupling, stationary field fluctuations, and a Markov--secular approximation for the reduced probe dynamics~\cite{campaioli2024quantum}. The resulting master equation takes the form
\begin{equation}
    \dot\rho_{\mathrm p}
    =
    -\frac{i}{\hbar}
    \comm{\hat H_{\mathrm p}+\hat H_{\mathrm{LS}}}{\rho_{\mathrm p}}
    +
    \Gamma_{\downarrow}(\Omega)\mathcal D _ {\hat c_{\mathrm p}}[\rho_{\mathrm p}]
    +
    \Gamma_{\uparrow}(\Omega)\mathcal D _ {\hat c_{\mathrm p}^\dagger}[\rho_{\mathrm p}],
    \label{eq:probe_master_equation}
\end{equation}
where the Lindbladian is defined as $\mathcal D _ {\hat L}[\rho]:=\hat L\rho\hat L^\dagger-\frac12\{\hat L^\dagger\hat L,\rho\}$. The Lamb shift is $\hat H_{\mathrm{LS}}=\hbar\delta\Omega\hat c_{\mathrm p}^\dagger\hat c_{\mathrm p}$. With the stationary unsymmetrised electric-field spectra defined by
\begin{subequations}
\label{eq:probe_field_spectra}
\begin{align}
    \mathcal S^{>}_{ij}(\omega)
    &=
    \int_{-\infty}^{\infty}
    \expval{\hat E_i^{-}(\vb r_{\mathrm p},\tau)
             \hat E_j^{+}(\vb r_{\mathrm p},0)}
    e^{i\omega\tau}
    \dd{\tau},\\
    \mathcal S^{<}_{ij}(\omega)
    &=
    \int_{-\infty}^{\infty}
    \expval{\hat E_i^{+}(\vb r_{\mathrm p},0)
             \hat E_j^{-}(\vb r_{\mathrm p},\tau)}
    e^{i\omega\tau}
    \dd{\tau},
\end{align}
\end{subequations}
the frequency-resolved transition rates are
\begin{align}
  \Gamma_{\downarrow}(\omega)
  =
  \frac{
    \vb d_{\mathrm p}^{\,*}
  }{\hbar}
  \cdot
  \boldsymbol{\mathcal S}^{>}(\omega)
  \cdot
  \frac{
    \vb d_{\mathrm p}
  }{\hbar},
  \quad
  \Gamma_{\uparrow}(\omega)
  =
  \frac{
    \vb d_{\mathrm p}
  }{\hbar}
  \cdot
  \boldsymbol{\mathcal S}^{<}(\omega)
  \cdot
  \frac{
    \vb d_{\mathrm p}^{\,*}
  }{\hbar},
  \label{eq:probe_transition_rates}
\end{align}
and the Lamb shift is
\begin{equation}
  \delta\Omega
  =\operatorname{PV}
  \int_0^\infty
  \frac{\Gamma _ \downarrow(\omega)-\Gamma _ \uparrow(\omega)}{\Omega-\omega}
  \frac{\dd{\omega}}{2\pi}.
  \label{eq:probe_lamb_shift}
\end{equation}

For the contribution generated by the medium's noise current, the Green-tensor solution in Eq.~\eqref{eq:mqed_green_solution} expresses both spectra as spatial integrals over the pumped region. In particular, the excitation spectrum contains the microscopic factor $p_1\boldsymbol{\chi}''/(p_0-p_1)$, whose finite limit at transparency was established above.

The probe occupation $N_{\mathrm p}=\expval{\hat c_{\mathrm p}^\dagger\hat c_{\mathrm p}}$ consequently obeys (with rates evaluated at $\Omega$)
\begin{equation}
    \dv{N_{\mathrm p}}{t}
    =
    -(\Gamma_{\downarrow}-\Gamma_{\uparrow})N_{\mathrm p}
    +
    \Gamma_{\uparrow}.
    \label{eq:probe_occupation_dynamics}
\end{equation}
For the medium-generated field considered here, transparency ($p_0=p_1$) gives $\Gamma_{\downarrow}=\Gamma_{\uparrow}>0$, even though the dissipative susceptibility vanishes. This equality refers to the medium-induced rates; any independent vacuum-field contribution is not included. For a probe initially in its ground state, the initial heating rate is then $\dot N_{\mathrm p}(0)=\Gamma_{\uparrow}>0$. This statement does not require a finite stationary probe occupation: in the absence of additional damping, the linear Markovian model predicts continued growth, while other loss channels or nonlinear effects determine the long-time behaviour.

The probe thus distinguishes a medium with vanishing linear absorption and finite fluctuations from a genuinely noiseless environment. More broadly, it demonstrates why the macroscopic susceptibility alone is insufficient to specify the quantum fluctuations of a pumped medium: the microscopic populations enter the excitation and relaxation spectra separately, even when their difference, and hence the net absorptive response, is zero.

\section{Conclusion}
\label{sec:conclusion}

We have developed a microscopic foundation for macroscopic quantum electrodynamics in optically pumped media.
Starting from a driven three-level system, we derived collective Langevin dynamics and established the corresponding electromagnetic susceptibility and noise-current correlations within a unified framework.

Our results demonstrate that the optical response and quantum fluctuations are governed by distinct combinations of the stationary populations.
In particular, the susceptibility depends on the population difference, whereas the noise correlations retain separate contributions from the two populations.
This distinction becomes especially important at transparency, where the dissipative susceptibility vanishes while the noise remains finite.
We further showed that these fluctuations can induce finite heating of a nearby quantum probe.

The framework thus provides a microscopic interpretation of the reservoir degrees of freedom employed in macroscopic QED and clarifies the relationship between optical gain, dissipation, and quantum fluctuations in nonequilibrium electromagnetic media.

\begin{acknowledgments}
  D.O.~is supported by the RIKEN Special Postdoctoral Researcher Program.
  This work was supported by JST PRESTO (No.~JPMJPR2693), JSPS KAKENHI (Grants No.~JP26K17783 and No.~JP26H00686), and the RIKEN Incentive Research Project.
\end{acknowledgments}

\section*{Data availability}
The data that support the findings of this article are not publicly available.
The data are available upon reasonable request from the authors.

\appendix
\section{Projection onto the effective two-level dynamics}
\label{appx:effective_lindblad}

We make explicit the elimination of the auxiliary level underlying Eq.~\eqref{eq:effective_lindblad}. In the pump-rotating frame, the three-level Hamiltonian is given by Eq.~\eqref{eq:three_level_hamiltonian}. Following the Born--Markov treatment of the reservoirs described in Sec.~\ref{sec:microscopic_model}, we write the microscopic master equation as
\begin{equation}
 \dot\rho=\mathcal L\rho=-\frac{i}{\hbar}\comm{H}{\rho}
 +\sum_{(i,j)}\gamma_{ij}\mathcal D _ {\ketbra{i}{j}}[\rho]
 +\mathcal L_{\phi}[\rho],
 \label{eq:appx_three_level_liouvillian}
\end{equation}
where the sum includes the transitions $\ket{2}\leftrightarrow\ket{1}$ and $\ket{1}\leftrightarrow\ket{0}$, with $\gamma_{21}=d_{21}$, $\gamma_{12}=u_{21}$, $\gamma_{10}=d_{10}$, and $\gamma_{01}=u_{10}$. Here $\mathcal L_{\phi}$ denotes possible population-preserving dephasing channels. The coherent pump is retained in $H$ rather than represented as a jump process.

Let $\Pi=\ketbra{0}{0}+\ketbra{1}{1}$ and define the Liouville-space projections
\begin{equation}
 \mathcal P X=\Pi X\Pi,\qquad
 \mathcal Q X=X-\Pi X\Pi.
 \label{eq:appx_explicit_projections}
\end{equation}
In the ordered basis $\{\ket{0},\ket{1},\ket{2}\}$, these act on an arbitrary operator as
\begin{equation}
 \mathcal P X=\begin{pmatrix}
 X_{00}&X_{01}&0\\
 X_{10}&X_{11}&0\\
 0&0&0
 \end{pmatrix},\quad
 \mathcal Q X=\begin{pmatrix}
 0&0&X_{02}\\
 0&0&X_{12}\\
 X_{20}&X_{21}&X_{22}
 \end{pmatrix}.
 \label{eq:appx_projection_matrices}
\end{equation}
Thus $\mathcal P$ retains the effective 2LS populations and coherence, while $\mathcal Q$ contains the auxiliary population and coherences involving $\ket{2}$. These are superoperator projections, not a decomposition into two independent Hilbert-space subsystems.

The projected equations are exactly
\begin{align}
 \dv{}{t}\mathcal P\rho
 &=\mathcal P\mathcal L\mathcal P\rho
   +\mathcal P\mathcal L\mathcal Q\rho,\\
 \dv{}{t}\mathcal Q\rho
 &=\mathcal Q\mathcal L\mathcal P\rho
   +\mathcal Q\mathcal L\mathcal Q\rho.
 \label{eq:appx_projected_equations}
\end{align}
When the $\mathcal Q$-sector relaxes rapidly and its initial transient has decayed, setting its time derivative to zero gives, formally,
\begin{equation}
 \mathcal L_{\mathrm{slow}}
 \simeq\mathcal P\mathcal L\mathcal P
 -\mathcal P\mathcal L\mathcal Q
 (\mathcal Q\mathcal L\mathcal Q)^{-1}
 \mathcal Q\mathcal L\mathcal P.
 \label{eq:appx_effective_generator}
\end{equation}
The inverse is restricted to the decaying fast subspace. Equation~\eqref{eq:appx_effective_generator} is a time-scale-separation formula; by itself it does not guarantee a trace-preserving Lindblad generator on the retained sector.

To identify the population-transfer terms explicitly, we first eliminate the rapidly decaying pump coherence $\rho_{02}$. As derived in Appendix~\ref{appx:rate_eq}, this produces the bidirectional rate $W$ in Eq.~\eqref{eq:pump_rate}. The resulting auxiliary-level population $p _ 2 := \expval{\rho}{2}$ obeys
\begin{equation}
 \dot p_2=Wp_0+u_{21}p_1-(W+d_{21})p_2.
\end{equation}
Remind that the excited-state population is $p _ 1 := \expval{\rho}{1}$.
In the regime $W+d_{21}$ is large compared with the effective 2LS rates, $\dot p_2\simeq0$ yields Eq.~\eqref{eq:p2_adiabatic}. Substitution into the excited-state population equation then gives
\begin{equation}
 \dot p_1=R_\uparrow p_0-R_\downarrow p_1,
\end{equation}
with the explicit $R_\uparrow$ and $R_\downarrow$ of Eq.~\eqref{eq:effective_rates}. The calculation of these coefficients is given in Appendix~\ref{appx:rate_eq} and is not repeated here.

For the effective 2LS, the population transfers are represented by the jump operators $\sigma_+=\ketbra{1}{0}$ and $\sigma_-=\ketbra{0}{1}$. Following the standard procedure in the theory of quantum open system~\cite{manzano2020short,campaioli2024quantum}, after neglecting nonsecular terms and incorporating the frequency renormalisation into $H_{\mathrm{eff}}$, the resulting generator takes the form
\begin{equation}
 \begin{aligned}
 \mathcal L_{\mathrm{2LS}}\rho={}&-\frac{i}{\hbar}\comm{H_{\mathrm{eff}}}{\rho}
 +R_\uparrow\mathcal D_{\sigma_+}[\rho]\\
 &+R_\downarrow\mathcal D_{\sigma_-}[\rho]
 +\frac{\gamma_\phi}{2}\mathcal D_{\sigma_z}[\rho],
 \end{aligned}
 \label{eq:appx_reduced_lindblad}
\end{equation}
which is Eq.~\eqref{eq:effective_lindblad}. A longitudinal coupling of the effective 2LS to a short-correlated reservoir provides, for example, a microscopic mechanism for the last term. With the dissipator convention used here, its contribution to coherence decay is $\gamma_\phi$, so that $\Gamma_2=(R_\uparrow+R_\downarrow)/2+\gamma_\phi$, as in Eq.~\eqref{eq:Gamma2}.

Finally, $\mathcal P$ is not trace preserving: $\Tr(\mathcal P\rho)=1-p_2$. Accordingly, interpreting the reduced state as a normalised 2LS density operator requires $p_2\ll1$. This is discussed further in Appendix~\ref{appx:rate_eq}.

\section{Pump-induced transitions and population rate equations}
\label{appx:rate_eq}

We derive the effective incoherent pump rate appearing in Eq.~\eqref{eq:pump_rate}. For the coherently driven $\ket{0}\leftrightarrow\ket{2}$ transition, let $\rho_{02} := \mel{0}{\rho}{2}$ be the coherence in the pump-rotating frame. With the Hamiltonian convention of Eq.~\eqref{eq:three_level_hamiltonian} and transverse decay rate $\kappa_{02}=(d_{21}+u_{10})/2 +\kappa_\phi$, its equation reads
\begin{equation}
 \dot\rho_{02}=(i\Delta-\kappa_{02})\rho_{02}
 -\frac{ig}{2}(p_2-p_0).
\end{equation}
Adiabatic elimination of this rapidly relaxing coherence gives
\begin{equation}
 \rho_{02}\simeq\frac{ig(p_0-p_2)}{2(\kappa_{02}-i\Delta)}.
\end{equation}
The coherent contribution to the upper-level population is consequently
\begin{equation}
 \left.\dot p_2\right|_{\mathrm{pump}}
 =W(p_0-p_2),\qquad
 W=\frac{g^2\kappa_{02}}{2(\Delta^2+\kappa_{02}^2)}.
 \label{eq:appx_pump_balance}
\end{equation}
This establishes the bidirectional nature of the effective pump transition.

Including the incoherent transitions between the three levels, the population equations are
\begin{align}
 \dot p_2&=W(p_0-p_2)+u_{21}p_1-d_{21}p_2,\nonumber\\
 \dot p_1&=d_{21}p_2+u_{10}p_0
 -(u_{21}+d_{10})p_1,\nonumber\\
 \dot p_0&=-W(p_0-p_2)+d_{10}p_1-u_{10}p_0.
 \label{eq:appx_three_level_rates}
\end{align}
For $\dot p_2\simeq0$ one obtains Eq.~\eqref{eq:p2_adiabatic}. Substitution into $\dot p_1$ yields
\begin{equation}
 \dot p_1=
 \left(u_{10}+\frac{Wd_{21}}{W+d_{21}}\right)p_0
 -\left(d_{10}+\frac{Wu_{21}}{W+d_{21}}\right)p_1,
 \label{eq:appx_effective_population}
\end{equation}
which reproduces the rates in Eq.~\eqref{eq:effective_rates}.

An important qualification is that $p_0+p_1=1-p_2$ in the full three-level model. The effective 2LS normalisation $p_0+p_1=1$ therefore additionally requires $p_2\ll1$. The validity of the approximation depends on the relative pump, relaxation, and coherence-decay timescales. In particular, the rate reduction alone does not justify neglecting a substantially occupied auxiliary level.

\section{Quantum central limit of collective 2LS fluctuations}
\label{appx:qclt}

Consider $N$ independent 2LSs prepared in the stationary product state $\rho_{\mathrm{ss}}^{\otimes N}$ of Eq.~\eqref{eq:steady_state_density}. Define the centred collective operators
\begin{equation}
 F_-^{(N)}=\frac{1}{\sqrt N}\sum_{\alpha=1}^N\sigma_-^{(\alpha)},
 \qquad F_+^{(N)}=\qty(F_-^{(N)})^\dagger.
\end{equation}
The stationary coherences vanish, so no further subtraction is needed. Independence implies
\begin{align}
 \expval{F_-^{(N)}F_+^{(N)}}&=p_0,\\
 \expval{F_+^{(N)}F_-^{(N)}}&=p_1.
\end{align}
Moreover,
\begin{equation}
 \comm{F_-^{(N)}}{F_+^{(N)}}
 =-\frac1N\sum_{\alpha=1}^N\sigma_z^{(\alpha)}.
\end{equation}
By the law of large numbers, the right-hand side converges in the stationary fluctuation representation to $p_0-p_1$. Thus the limiting collective fluctuation algebra is
\begin{equation}
 \comm{F_-}{F_+}=p_0-p_1,
 \label{eq:appx_collective_commutator}
\end{equation}
with the above second moments.

To see why the limiting state is Gaussian, expand any joint cumulant of order $m$ of the centred sums. Independence eliminates terms whose site labels do not form a connected contribution; the surviving terms scale as $N^{1-m/2}$. Consequently, all cumulants of order $m>2$ vanish as $N\to\infty$, while the second cumulants remain finite. The quantum central limit therefore produces a quasifree fluctuation state.

One explicit representation uses two independent canonical bosonic modes $a,b$ in their joint vacuum:
\begin{equation}
 F_-=\sqrt{p_0}\,a+\sqrt{p_1}\,b^\dagger,
 \qquad F_+=F_-^\dagger.
 \label{eq:appx_bosonic_limit}
\end{equation}
It reproduces both the commutator and the two stationary contractions. This two-mode representation remains regular at $p_0=p_1$, where the limiting commutator vanishes but the fluctuation variance does not. It is a representation of the limiting fluctuation algebra, not an identification of the microscopic 2LS Hilbert space with two oscillator Hilbert spaces.

\section{Bosonic dilation of the collective Langevin dynamics}
\label{appx:langevin}

We verify the Langevin representation in Sec.~\ref{sec:langevin} using two independent damped bosonic modes. Let $a,b$ obey
\begin{align}
 \dot a&=-(i\widetilde\omega_0+\Gamma_2)a
 +\sqrt{2\Gamma_2}\,a_{\mathrm{in}},\\
 \dot b^\dagger&=-(i\widetilde\omega_0+\Gamma_2)b^\dagger
 +\sqrt{2\Gamma_2}\,b_{\mathrm{in}}^\dagger.
\end{align}
The second equation is the adjoint of the equation for a mode $b$ with the opposite free rotation. Both inputs are independent vacuum white-noise fields, and both oscillator modes have damping rate $\Gamma_2$.

Define the collective fluctuation
\begin{equation}
 F_-=\sqrt{p_0}\,a+\sqrt{p_1}\,b^\dagger.
\end{equation}
It follows directly that
\begin{equation}
 \dot F_-=-(i\widetilde\omega_0+\Gamma_2)F_-+\xi,
\end{equation}
where
\begin{equation}
 \xi=\sqrt{2\Gamma_2}
 \qty(\sqrt{p_0}\,a_{\mathrm{in}}
 +\sqrt{p_1}\,b_{\mathrm{in}}^\dagger).
\end{equation}
The vacuum input contractions give the two noise correlations in Eq.~\eqref{eq:langevin_noise_correlations}.

For a stationary solution with the initial transient removed,
\begin{equation}
 F_-(t)=\int_{-\infty}^{t}
 e^{-(i\widetilde\omega_0+\Gamma_2)(t-s)}\xi(s)\dd{s}.
\end{equation}
The white-noise correlations yield
\begin{align}
 \expval{F_-(t)F_+(t)}&=p_0,\\
 \expval{F_+(t)F_-(t)}&=p_1,
\end{align}
in agreement with the quantum central limit. Likewise, the equal-time commutator is $p_0-p_1$.

To check the stationary two-time correlations against the effective 2LS
Lindblad dynamics, we use the adjoint generator of
Eq.~\eqref{eq:effective_lindblad}. Its action on the coherence operators is
\begin{align}
 \mathcal L_{\mathrm{2LS}}^\dagger\sigma_-
 &=-(i\widetilde\omega_0+\Gamma_2)\sigma_-,\\
 \mathcal L_{\mathrm{2LS}}^\dagger\sigma_+
 &=(i\widetilde\omega_0-\Gamma_2)\sigma_+.
\end{align}
The quantum regression theorem therefore gives, for $\tau\geq0$,
\begin{align}
 \expval{\sigma_-(t+\tau)\sigma_+(t)}_{\mathrm{ss}}
 &=p_0e^{-(i\widetilde\omega_0+\Gamma_2)\tau},\\
 \expval{\sigma_+(t+\tau)\sigma_-(t)}_{\mathrm{ss}}
 &=p_1e^{(i\widetilde\omega_0-\Gamma_2)\tau}.
\end{align}
The Langevin solution above yields exactly the same expressions with
$\sigma_\mp$ replaced by $F_\mp$: the vacuum inputs at times later than
$t$ have vanishing contractions with the stationary operators at $t$,
while the equal-time correlations supply the factors $p_0$ and $p_1$.
Thus the bosonic representation reproduces the stationary two-time
coherence correlations of the effective 2LS Lindblad dynamics.

\section{Isotropic media from randomly oriented transition dipoles}
\label{appx:isotropic}

The main text assumes a fixed transition dipole $\vb d$. We now consider $N$ independent 2LSs in each mesoscopic cell with dipoles $\vb d_\alpha=|\vb d|\vb u_\alpha$, where $\vb u_\alpha$ are independently distributed unit vectors with isotropic second moment
\begin{equation}
 \overline{u_{\alpha i}u_{\alpha j}^*}=\frac{\delta_{ij}}{3}.
 \label{eq:appx_orientation_average}
\end{equation}
For definiteness, the orientations are fixed during the fluctuation dynamics, and all 2LSs share the same stationary populations and linewidth. Their microscopic fluctuations remain independent before orientational coarse-graining.

The local susceptibility is the sum of the rank-one contributions. Self-averaging as $N\to\infty$ gives
\begin{equation}
 \chi_{ij}(\omega)=
 \frac{n|\vb d|^2}{3\epsilon_0\hbar}
 \frac{p_0-p_1}{\widetilde\omega_0-\omega-i\Gamma_2}
 \delta_{ij}.
 \label{eq:appx_isotropic_chi}
\end{equation}
The corresponding noise polarisation must be coarse-grained from the independent microscopic noises, rather than inferred solely from the averaged susceptibility.

Let $a_\alpha,b_\alpha$ denote independent canonical vacuum channels for each 2LS. Define the collective Cartesian channels
\begin{equation}
 A_i=\sqrt{\frac3N}\sum_{\alpha=1}^N
 u_{\alpha i}a_\alpha,\qquad
 B_i=\sqrt{\frac3N}\sum_{\alpha=1}^N
 u_{\alpha i}^*b_\alpha.
 \label{eq:appx_isotropic_channels}
\end{equation}
In the self-averaging limit, their commutators tend to
\begin{equation}
 \comm{A_i}{A_j^\dagger}
 =\comm{B_i}{B_j^\dagger}=\delta_{ij},
\end{equation}
with vanishing mixed commutators. The complex conjugation in the definition of $B_i$ ensures that the creation contribution carries the same dipole orientation as the annihilation contribution. The collective vector fluctuation can consequently be represented as
\begin{equation}
 \vb F=\sqrt{\frac N3}
 \qty(\sqrt{p_0}\,\vb A+\sqrt{p_1}\,\vb B^\dagger),
\end{equation}
where the equality is understood in the limiting Gaussian fluctuation representation. In particular, the two unsymmetrised correlations are proportional to $p_0\delta_{ij}$ and $p_1\delta_{ij}$, respectively.

In the absorptive regime, three effective canonical noise operators may be introduced componentwise,
\begin{equation}
 c_i=\sqrt{\frac{p_0}{p_0-p_1}}A_i
 +\sqrt{\frac{p_1}{p_0-p_1}}B_i^\dagger,
 \label{eq:appx_isotropic_effective_modes}
\end{equation}
so that $\comm{c_i}{c_j^\dagger}=\delta_{ij}$ and $\expval{c_i^\dagger c_j}=p_1\delta_{ij}/(p_0-p_1)$. The gain regime follows by exchanging annihilation and creation roles, as in Appendix~\ref{appx:mqed_consistency}. At transparency, the canonical rescaling becomes singular, while the original $A_i,B_i$ noise representation stays regular.

The emergence of three effective Cartesian channels is a consequence of isotropic orientational averaging and the independence of the microscopic fluctuations.

\section{Consistency with conventional macroscopic QED}
\label{appx:mqed_consistency}

In this appendix, we make explicit the Fourier normalisation and operator transformations underlying the correspondence between the microscopic Langevin noise in Sec.~\ref{sec:langevin} and the conventional macroscopic-QED noise polarisation in Sec.~\ref{sec:mqed}. We retain the fixed dipole orientation $\vb u=\vb d/|\vb d|$ and the positive-frequency convention used in the main text.

\subsection{Fourier normalisation and the Langevin phase}

The independent vacuum inputs in Eq.~\eqref{eq:langevin_noise_representation} obey
\begin{equation}
    \comm{\hat a_{\mathrm{in}}(\vb r,t)}{\hat a_{\mathrm{in}}^\dagger(\vb r',t')}
    =\comm{\hat b_{\mathrm{in}}(\vb r,t)}{\hat b_{\mathrm{in}}^\dagger(\vb r',t')}
    =\delta_{\vb r,\vb r'}\delta_{t,t'},
\end{equation}
with vanishing cross-commutators. Here the inputs are continuum normalised in space, as in Eq.~\eqref{eq:continuum_noise_correlations}. We introduce frequency-domain operators by
\begin{equation}
\begin{aligned}
    \hat a_0(\vb r,\omega)
    &=\frac{1}{\sqrt{2\pi}}\int_{-\infty}^{\infty}
    \hat a_{\mathrm{in}}(\vb r,t)e^{i\omega t}\dd{t},\\
    \hat b_0^\dagger(\vb r,\omega)
    &=\frac{1}{\sqrt{2\pi}}\int_{-\infty}^{\infty}
    \hat b_{\mathrm{in}}^\dagger(\vb r,t)e^{i\omega t}\dd{t}.
\end{aligned}
\label{eq:appx_frequency_inputs}
\end{equation}
Thus $\hat b_0^\dagger(\vb r,\omega)$ denotes the Fourier component of the creation input at frequency $\omega$; equivalently it is the adjoint of the annihilation input at frequency $-\omega$. These operators satisfy
\begin{equation}
\begin{aligned}
    \comm{\hat a_0(\vb r,\omega)}{\hat a_0^\dagger(\vb r',\omega')}
    &=\delta_{\vb r,\vb r'}\delta_{\omega,\omega'},\\
    \comm{\hat b_0(\vb r,\omega)}{\hat b_0^\dagger(\vb r',\omega')}
    &=\delta_{\vb r,\vb r'}\delta_{\omega,\omega'},
\end{aligned}
\end{equation}
where $\hat b_0(\vb r,\omega)$ is the adjoint of $\hat b_0^\dagger(\vb r,\omega)$. The vacuum contractions have the same delta normalisation.

Using the MQED frequency convention in Eq.~\eqref{eq:mqed_fourier_convention}, the continuum Langevin force has the positive-frequency component
\begin{equation}
    \hat\zeta^-(\vb r,\omega)
    =\sqrt{\frac{\Gamma_2}{\pi}}
    \qty[\sqrt{p_0}\,\hat a_0(\vb r,\omega)
    +\sqrt{p_1}\,\hat b_0^\dagger(\vb r,\omega)].
    \label{eq:appx_zeta_frequency}
\end{equation}
The noise polarisation in Eq.~\eqref{eq:noise_polarisation} is consequently
\begin{equation}
\begin{aligned}
    \hat{\vb P}_{\mathrm N}^-(\vb r,\omega)
    &=\vb d\sqrt{\frac{n\Gamma_2}{\pi}}
    \frac{1}{\Gamma_2+i\Delta _ \omega}
    \qty[\sqrt{p_0}\,\hat a_0
    +\sqrt{p_1}\,\hat b_0^\dagger],
\end{aligned}
    \label{eq:appx_noise_raw}
\end{equation}
where $\Delta _ \omega=\widetilde\omega_0-\omega$ and the arguments of the input operators have been suppressed. Writing
\begin{equation}
    \frac{1}{\Gamma_2+i\Delta _ \omega}
    =\frac{e^{i\phi(\omega)}}{\sqrt{\Gamma_2^2+\Delta _ \omega^2}},
    \quad
    e^{i\phi(\omega)}
    =\frac{\Gamma_2-i\Delta _ \omega}{\sqrt{\Gamma_2^2+\Delta _ \omega^2}},
    \label{eq:appx_lorentzian_phase}
\end{equation}
we absorb the common phase into the canonical inputs according to
\begin{equation}
    \hat a=e^{i\phi}\hat a_0,
    \quad
    \hat b^\dagger=e^{i\phi}\hat b_0^\dagger.
    \label{eq:appx_phase_rotation}
\end{equation}
Both transformations preserve the vacuum correlations and canonical commutators. The resulting expression is
\begin{equation}
    \hat{\vb P}_{\mathrm N}^-
    =\vb u\sqrt{\frac{n|\vb d|^2\Gamma_2}{\pi(\Gamma_2^2+\Delta _ \omega^2)}}
    \qty[\sqrt{p_0}\,\hat a+\sqrt{p_1}\,\hat b^\dagger].
    \label{eq:appx_noise_phase_absorbed}
\end{equation}

\subsection{Absorptive and amplifying regimes}

For $p_0>p_1$, define the dipole-aligned canonical operator
\begin{equation}
    \hat c_{\mathrm{abs}}
    =\sqrt{\frac{p_0}{p_0-p_1}}\,\hat a
    +\sqrt{\frac{p_1}{p_0-p_1}}\,\hat b^\dagger.
    \label{eq:appx_absorption_operator}
\end{equation}
It satisfies $\comm{\hat c_{\mathrm{abs}}(\vb r,\omega)}{\hat c_{\mathrm{abs}}^\dagger(\vb r',\omega')}=\delta_{\vb r,\vb r'}\delta_{\omega,\omega'}$, and Eq.~\eqref{eq:appx_noise_phase_absorbed} becomes
\begin{equation}
    \hat{\vb P}_{\mathrm N}^-
    =\sqrt{\frac{\hbar\epsilon_0}{\pi}}
    \sqrt{\boldsymbol\chi''}\cdot
    \qty(\vb u\hat c_{\mathrm{abs}}).
    \label{eq:appx_absorption_mqed}
\end{equation}
This is precisely the single-channel representation used in Eqs.~\eqref{eq:mqed_standard_noise} and \eqref{eq:mqed_effective_noise}.

For $p_1>p_0$, the scalar susceptibility eigenvalue $\chi''_\parallel$ is negative. We instead define
\begin{equation}
    \hat c_{\mathrm{amp}}
    =\sqrt{\frac{p_1}{p_1-p_0}}\,\hat b
    +\sqrt{\frac{p_0}{p_1-p_0}}\,\hat a^\dagger,
    \label{eq:appx_gain_operator}
\end{equation}
which is again canonical. Since $\hat c_{\mathrm{amp}}^\dagger$ reproduces the bracket in Eq.~\eqref{eq:appx_noise_phase_absorbed}, the noise polarisation takes the amplifying-medium form
\begin{equation}
    \hat{\vb P}_{\mathrm N}^-
    =\sqrt{\frac{\hbar\epsilon_0}{\pi}}
    \sqrt{-\boldsymbol\chi''}\cdot
    \qty(\vb u\hat c_{\mathrm{amp}}^\dagger).
    \label{eq:appx_gain_mqed}
\end{equation}
The square roots in Eqs.~\eqref{eq:appx_absorption_mqed} and \eqref{eq:appx_gain_mqed} act on the respective positive-semidefinite rank-one tensors. 

\subsection{Transparency limit}

At $p_0=p_1$, neither $\hat c_{\mathrm{abs}}$ nor $\hat c_{\mathrm{amp}}$ has a finite normalisation, because the commutator of the unnormalised bracket in Eq.~\eqref{eq:appx_noise_phase_absorbed} vanishes. Nevertheless, Eq.~\eqref{eq:appx_noise_phase_absorbed} itself is regular: its prefactor and both vacuum-noise contributions remain finite. In particular,
\begin{equation}
\begin{aligned}
    \expval{\hat P_{\mathrm N,i}^-(\vb r,\omega)
    \hat P_{\mathrm N,j}^+(\vb r',\omega')}
    &=\frac{n\Gamma_2}{\pi(\Gamma_2^2+\Delta _ \omega^2)}
    p_0 d_i d_j^*\,\delta_{\vb r,\vb r'}\delta_{\omega,\omega'},\\
    \expval{\hat P_{\mathrm N,i}^+(\vb r,\omega)
    \hat P_{\mathrm N,j}^-(\vb r',\omega')}
    &=\frac{n\Gamma_2}{\pi(\Gamma_2^2+\Delta _ \omega^2)}
    p_1 d_j d_i^*\,\delta_{\vb r,\vb r'}\delta_{\omega,\omega'}.
\end{aligned}
    \label{eq:appx_transparency_correlations}
\end{equation}
These expressions remain finite at $p_0=p_1$, while their difference, and hence the noise-polarisation commutator, vanishes. Thus the apparent singularity of the conventional one-canonical-mode parametrisation at transparency reflects a change in the operator normalisation, not a divergence or disappearance of the microscopic noise.

\bibliography{%
  bib/all,%
  textbook/textbook_all%
}
\end{document}